\documentclass[
reprint,
amssymb, amsmath,
aps,
frontmatterverbose,
longbibliography,
]{revtex4-1}

\usepackage{amsmath}
\usepackage{graphicx}
\usepackage{docs}%
\usepackage{bm}%
\usepackage{mathtools}
\usepackage[linkcolor=blue]{hyperref}%
\expandafter\ifx\csname package@font\endcsname\relax\else
\expandafter\expandafter
\expandafter\usepackage
\expandafter\expandafter
\expandafter{\csname package@font\endcsname}%
\fi
 \usepackage{multirow}
 \usepackage{tabularew}
 \usepackage[flushleft]{threeparttable}
\usepackage{booktabs}
\newcolumntype{C}{>{$}c<{$}}
\AtBeginDocument{
\heavyrulewidth=.08em
\lightrulewidth=.05em
\cmidrulewidth=.03em
\belowrulesep=.65ex
\belowbottomsep=0pt
\aboverulesep=.4ex
\abovetopsep=0pt
\cmidrulesep=\doublerulesep
\cmidrulekern=.5em
\defaultaddspace=.5em
}
\usepackage{multirow}
\usepackage{siunitx}
 \graphicspath{{./pdf/}{./jpeg/}{./eps/}}

\begin{document}

\title{Stochastic Spiking Neural Networks Enabled by Magnetic Tunnel Junctions: From Nontelegraphic to Telegraphic Switching Regimes}%

\author{Chamika M. Liyanagedera}
\email{cliyanag@purdue.edu}
\affiliation{Purdue University, West Lafayette, IN 47906}
\author{Abhronil Sengupta}
\affiliation{Purdue University, West Lafayette, IN 47906}
\author{Akhilesh Jaiswal} 
\affiliation{Purdue University, West Lafayette, IN 47906}
\author{Kaushik Roy }%
\affiliation{Purdue University, West Lafayette, IN 47906}
\date{\today}%

\begin{abstract}
Stochastic spiking neural networks based on nanoelectronic spin devices can be a possible pathway to achieving ``brainlike'' compact and energy-efficient cognitive intelligence. The computational model attempt to exploit the intrinsic device stochasticity of nanoelectronic synaptic or neural components to perform learning or inference. However, there has been limited analysis on the scaling effect of stochastic spin devices and its impact on the operation of such stochastic networks at the system level. This work attempts to explore the design space and analyze the performance of nanomagnet-based stochastic neuromorphic computing architectures for magnets with different barrier heights. We illustrate how the underlying network architecture must be modified to account for the random telegraphic switching behavior displayed by magnets with low barrier heights as they are scaled into the superparamagnetic regime. We perform a device-to-system-level analysis on a deep neural-network architecture for a digit-recognition problem on the MNIST data set.
\end{abstract}

\maketitle

\section{Introduction}
Emulating the computational primitives of neural-network-based machine-learning approaches by the inherent device physics of nanoelectronic components has proven to be useful in reducing the area and energy requirements of the underlying hardware fabrics. To that effect, several post-CMOS technologies, like phase-change memories \cite{kuzum2011nanoelectronic}, Ag-Si devices \cite{jo2010nanoscale}, and spintronic devices \cite{sengupta2016vision} among others, have been shown to exhibit neural and synaptic functionalities at the intrinsic device level. In this work, we focus on spintronic technologies in particular due to the low current and energy requirements of such devices in comparison to traditional memristive technologies. 

While traditional neuromorphic computing models have been based on deterministic neural and synaptic primitives, recent effort has been directed towards adapting such computing schemes to stochastic models. This endeavor has been driven primarily by two factors. (1) Deterministic neural or synaptic models are characterized by multibit resolution. However, as device dimensions of nanoelectronic neurons or synapses are scaled down, they might lose the multibit resolution capacity. In addition, such devices are expected to exhibit increased stochasticity during the switching process. For instance, spintronic devices exhibit stochasticity due to thermal noise at nonzero temperatures. Consequently, computational models that leverage the
underlying device stochasticity have recently been explored. Information encoding over time due to probabilistic synaptic or neural updates also enables state compression of neural and synaptic units, thereby allowing them to be implemented by single-bit technologies. (2) The human brain, the main inspiration behind such neuromorphic computing models, is characterized by stochastic neural and synaptic units. As a matter of fact, neuroscience studies have indicated that cortical neurons generate spikes probabilistically over time \cite{moreno2014poisson}. Consequently, stochastic neural computing models can potentially enable ``brain-like" cognitive computing. In this work, we focus on stochastic neural inference in deep neural networks for typical pattern recognition tasks \cite{probabilistic}. However, the analysis can be easily extended to stochastic synaptic units \cite{srinivasan2016magnetic}, or even other unconventional computing platforms that require stochastic switching elements like Ising computing \cite{shim2016ising,sutton2017intrinsic} and Bayesian inference, among others.

Spintronic devices have recently found wide application in large-scale neurocomputing hardware owing to their scalability and low power requirements. Spin-torque memristors with magnetic domain walls have been shown to be a suitable candidate for implementing multilevel neurosynapses \cite{lequeux2016magnetic} and integrating and fire spiking neurons \cite{sengupta2016vision}.Another study demonstrated that the inherent magnetic dynamics of a magnetic tunnel junction (MTJ) can be used to emulate the functionality of biologically inspired leaky-integrate and fire-spiking neurons \cite{sengupta2016magnetic}.   In Ref. \cite{vincent2015spin}, spin-transfer-torque MTJs were used as stochastic binary synapses, where the stochastic effects of the devices are used to perform unsupervised learning. It was also demonstrated that MTJs can be used as binary elements to implement long-term short-term stochastic synapses to improve the learning efficiency of a neural network \cite{srinivasan2016magnetic}. A review of bioinspired neuromorphic computing platforms based on spintronic devices can be found in Ref. \cite{grollier2016spintronic}.

As mentioned previously, spintronic devices display a stochastic switching nature due to thermal noise. Given a particular duration of write current flowing through the device, a magnet exhibits a particular probability of switching during that corresponding write cycle. Consecutive write and read cycles can be used to generate an output pulse stream whose average value depends on the magnitude of the input stimulus. While stochastic neural networks based on spintronic devices have been explored previously \cite{probabilistic,srinivasan2017magnetic}, there has been limited analysis of the scaling effects of these devices. It is generally expected that, as the magnet dimensions scale down, the device would exhibit increased stochasticity. Furthermore, the operating current or voltage ranges required for operating such devices in the probabilistic regime would be reduced. However, as the scaling tends to the superparamagnetic regime, the magnets undergo random telegraphic switching with a low data-retention time, making the device practically volatile in nature. Utilizing such a device as a biased random generator requires a rethinking of the peripherals and the underlying network architecture since parallel read and write operations of the nanomagnets are then required. However, the adaptation of such low-energy superparamagnets as neural components comes at the expense of reduced error resiliency. This is the case mainly because the gradient or the rate of change of the switching characteristics of such magnets in response to the input current magnitude is extremely high. This article attempts to address the different schemes of operation of stochastic spiking neural networks (SNNs) for magnets in nontelegraphic to telegraphic regimes and analyze its associated energy-accuracy trade-offs at the system level.

\section{Magnetic Tunnel Junction as a Stochastic Switching Element}

\begin{figure}[t!]
\centering
\includegraphics[width=3.2in]{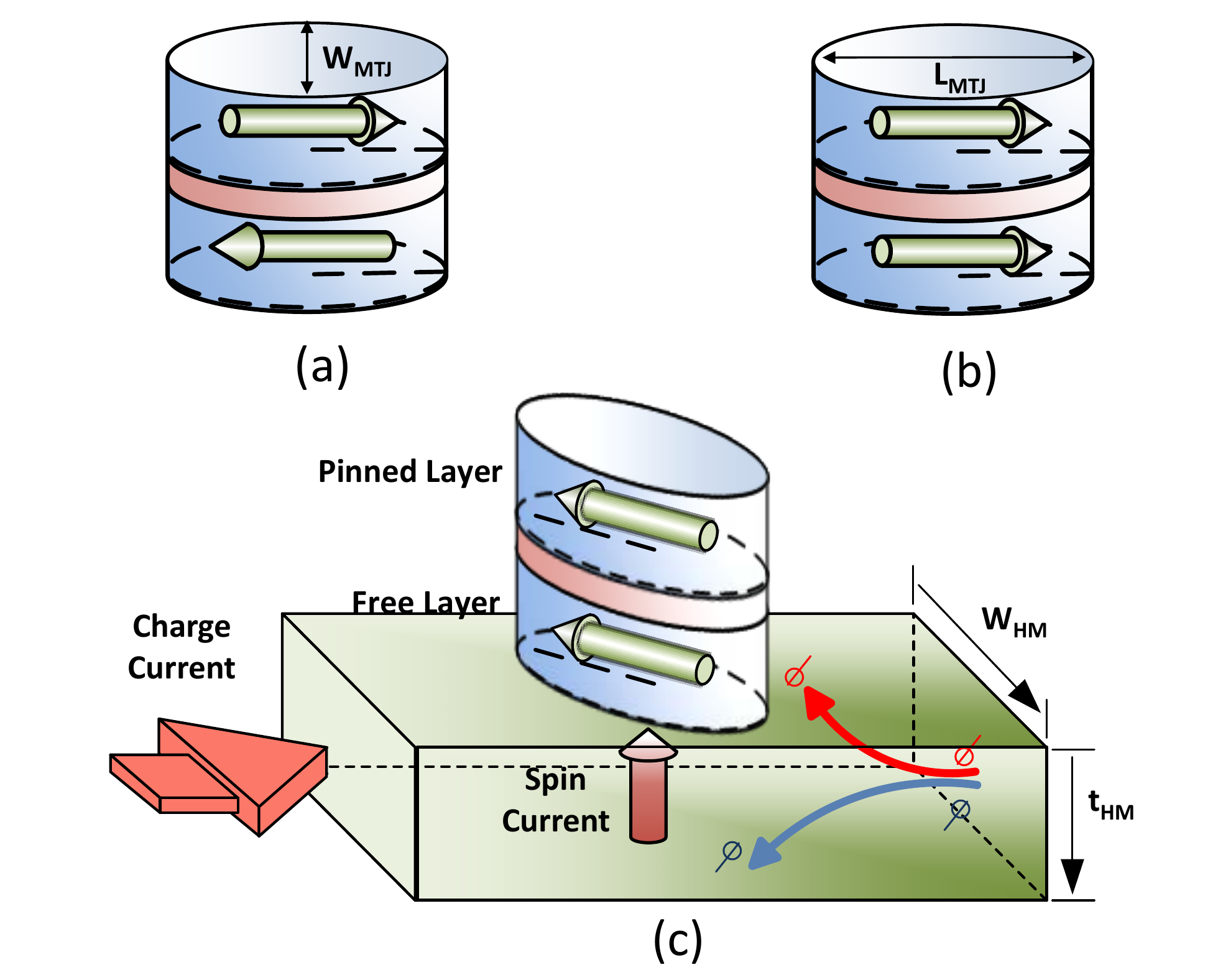}
\caption{(a) High resistive anti-parallel state of an MTJ, (b) Low resistive parallel state  of an MTJ, and (c) A SHE-MTJ device structure where the MTJ is switched by passing charge current through the underlying heavy metal. The charge current flowing through the heavy metal leads to spin splitting, thereby creating a perpendicular spin current, switching the magnetization direction of the free layer.}
\label{shemtj}
\end{figure}

A MTJ is a magnetoresistive device that consists of a tunneling oxide sandwiched between two magnetic contacts. One of the contacts is magnetically hardened and is called the \textit{pinned} layer, while the direction of magnetization of the other contact, called the  \textit{free} layer, can be switched. In a spin-Hall effect based MTJ (SHE-MTJ), the direction of the free layer is switched by passing a charge current through an underlying heavy metal (HM), as shown in Fig. \ref{shemtj}. The passage of the charge current ($I_{charge}$) through the HM layer induces a resulting spin current ($I_{spin}$) flowing perpendicular to the planes of the magnetic layers of the MTJ. This spin current can switch the direction of magnetization of the free layer, making it parallel (P) or anti-parallel (AP) to that of the pinned layer, through the well known spin-orbit torque mechanism \cite{liu2012spin,pai2012spin}. Owing to the magnetoresistance effect, the SHE-MTJ exhibits a lower resistance ($R_P$), when in the P state and a higher resistance ($R_{AP}$), when in the AP state. Thus, the SHE-MTJ shown in Fig. 2, exhibits decoupled read and write current paths. Write operation can be achieved by a charge current flowing through the HM layer, while the read operation can be accomplished by sensing the resistance of the MTJ in a direction transverse to the plane of the magnetic layers.

It is to be noted that the switching process of the nanoscale free layer is influenced by thermal noise at nonzero temperatures. Thermal noise results in a stochastic switching behavior wherein, for a given current flowing through the HM layer, the MTJ switches with a certain probability. Moreover, the probability of switching can be controlled by the magnitude of the current flowing through the HM. The dynamics of the magnetization vector in the presence of the HM-layer current is given by the stochastic Landau-Lifshitz-Gilbert-Slonczewski (LLGS) equation and
can be written as \cite{jaiswalspin}

\begin{equation}
\label{LLG_eqn}
\frac{\partial \widehat{m}}{\partial \tau} = - \widehat{m} \times \vec{H}_{eff} - \alpha \widehat{m} \times \widehat{m} \times \vec{H}_{eff}  + \frac{1}{\vert \gamma \vert} (\alpha\widehat{m} \times \vec{STT} + \vec{STT})
\end{equation}
where $\tau$ is $\frac{\vert \gamma \vert }{1 + \alpha^2}t$.
\\

Here, $\alpha$ is the Gilbert-damping constant, $\gamma$ is the gyromagnetic ratio, $\widehat{m}$ is the unit vector in the direction of the magnetization, $t$ is the simulation time and $H_{eff}$ is the effective magnetic field including the demagnetization field and the interface anisotropy field. A detailed description of the various fields included in $H_{eff}$ can be found in Ref. \cite{jaiswalspin}. $\vec{STT} $ in Eq. (\ref{LLG_eqn}) is the term representing the torque due to the SHE effect (modeled as a spin-transfer torque term) and can be written as follows \cite{boltzmann},

\begin{equation}
\label{STT_eqn}
\vec{STT} = \vert \gamma \vert \beta (\widehat{m}\times(\epsilon_{SHE} \widehat{m}\times \widehat{mp} )),~ \beta = \frac{\hbar J_{q}}{2e\mu_oM_St_{FL}}
\end{equation}

where $\widehat{mp}$ is the magnetization of the pinned layer (PL), $e$ is charge of an electron, $\mu_o$ is the permeability of vacuum, $\hbar$ is modified Planck's constant, $t_{FL}$ is the thickness of the free layer (FL), and $M_S$ is saturation magnetization. $J_{q}$ is the charge current density flowing through the heavy metal. $\epsilon_{SHE}$ is the spin-polarization efficiency (defined as the ratio of the spin current generated due to the charge current flowing through the HM layer) and can be written as \cite{manipspin},

\begin{figure}[t!]
\centering
\includegraphics[width=3in]{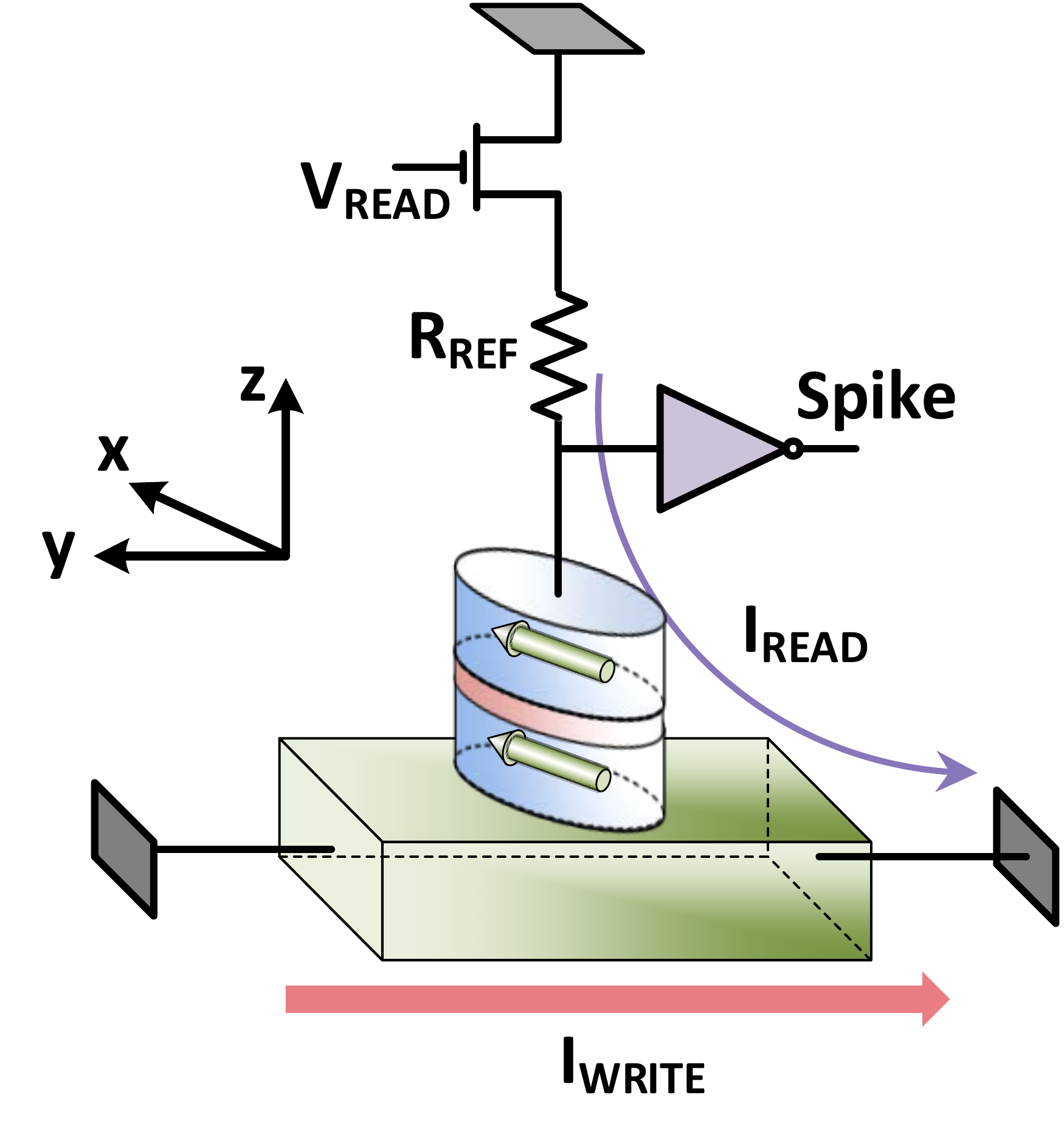}
\caption{\ Decoupled read and write current paths of the MTJ with HM. Output of the inverter will be high if the MTJ is in the P state, and low if the MTJ is in the AP state.  }
\label{shemtj2}
\end{figure}

\begin{equation}
\epsilon_{she}=   \frac{I_{spin}}{I_{charge}}=\frac{\pi w}{4t}\theta_{she}\left(1-sech\left(\frac{t}{\lambda_{sf}}\right)\right)
\end{equation}
where, $w$ is width of free layer, $t$ is thickness of heavy metal, $\theta_{she}$ is spin hall angle, $\lambda_{sf}$ is spin-flip length. 

The random switching process due to the effect of the thermal noise can be included in the LLGS equation through a stochastic field $\vec{H}_{thermal}$ in $\vec{H}_{eff}$ \cite{brownthermal}, 

\begin{figure}[t!]
\centering
\includegraphics[width=3.4in]{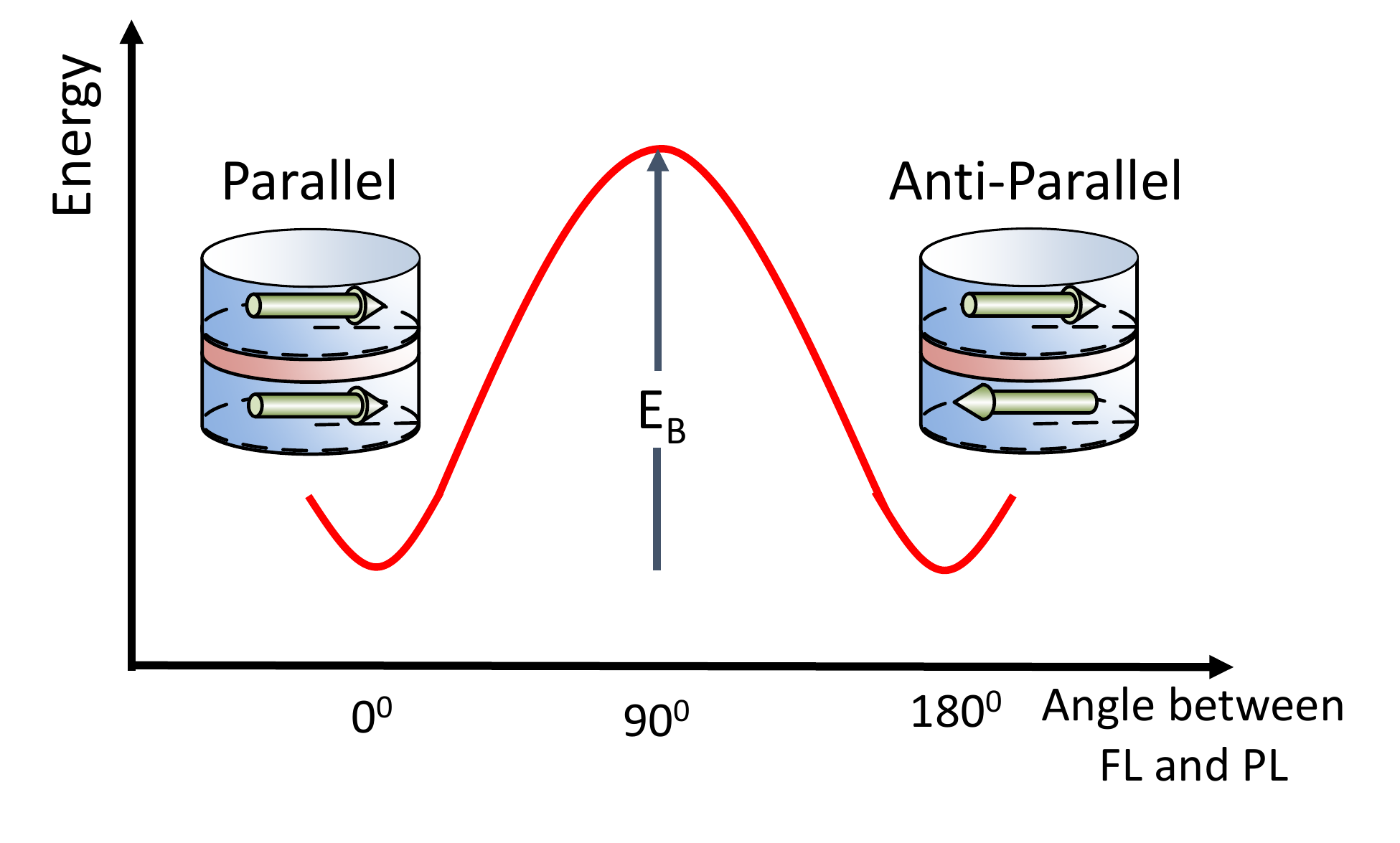}
\caption{\ The two operating states of an MTJ. The two states are thermally stable if the barrier height of the magnet, $E_B$, is large enough.}
\label{enbar}
\end{figure}

\begin{figure*}
\centering
\includegraphics[width=4.7in]{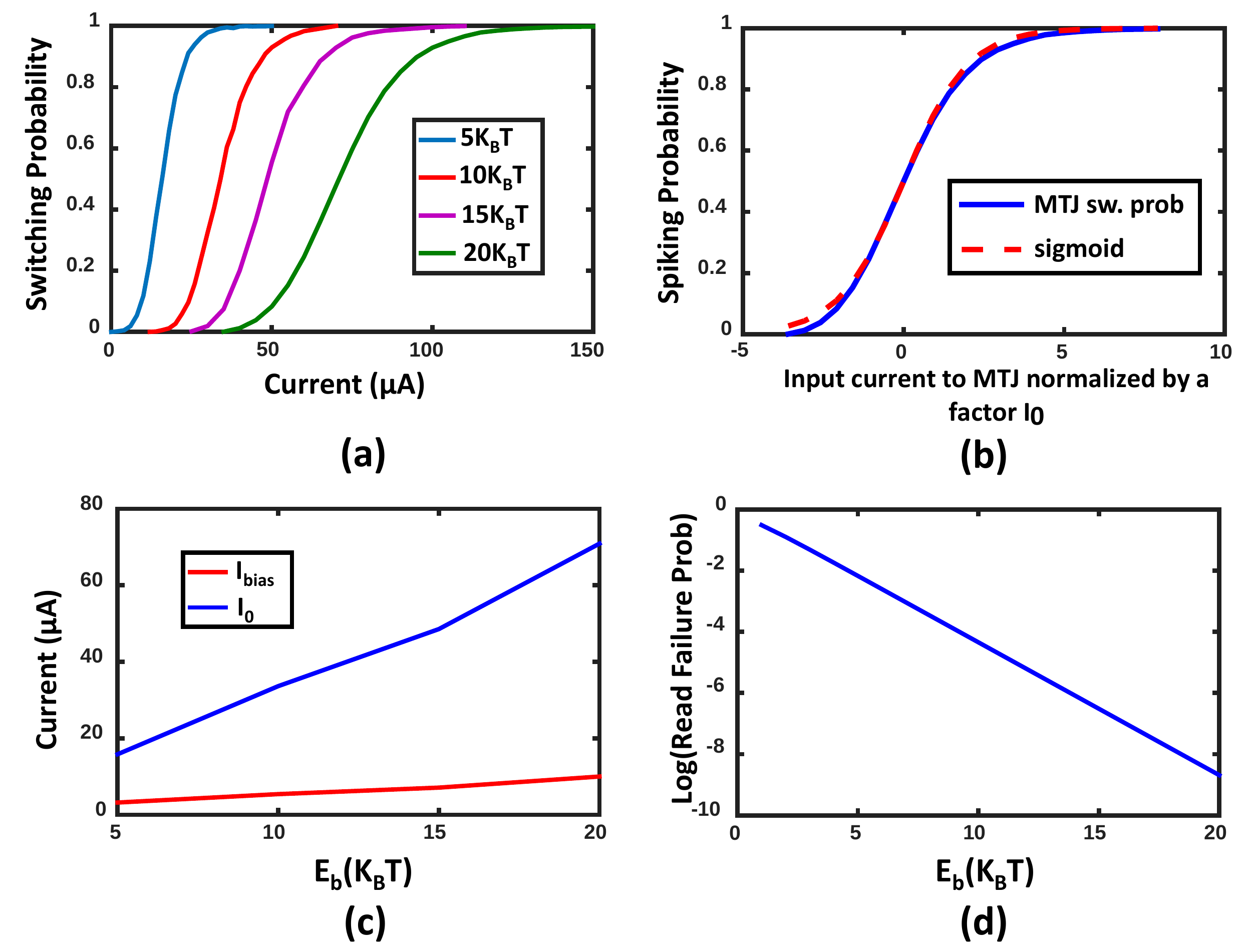}
\caption{\ (a) Switching characteristics of an MTJ with varying ${E}_{B}$ at $T=300K$ for a write cycle duration of $0.5ns$, (b) MTJ switching probability characteristics as a function of $I-{I}_{bias}$, normalized by a factor ${I}_{o}$. The data closely resembles the sigmoid function, (c) Variation of the bias current, ${I}_{bias}$, and the normalizing factor, ${I}_{o}$, with varying ${E}_{B}$. Both ${I}_{bias}$ and ${I}_{o}$ decrease with decreasing ${E}_{B}$, (d) Failure probability during a read cycle of $1ns$ (in logarithm scale) with varying $E_B$. }
\label{varyeb}
\end{figure*}

\begin{equation}
\label{therm_eqn}
\vec{H}_{thermal} = \vec{\zeta}\sqrt{\frac{2\alpha k_BT}{\vert \gamma \vert dt M_SVol}}
\end{equation}

where, $k_B$ is the Boltzmann constant, $T$ is the temperature, $Vol$ is volume of the free layer magnet and $dt$ is the simulation time step. The term $\vec{\zeta}$ in Eq. \ref{therm_eqn} is a Gaussian random variable with zero mean and a standard deviation equal to 1. The inclusion of thermal noise turns the LLG equations into a stochastic differential equation. We used Heun's method to integrate the stochastic LLG equation. The details of applying Heun's method to stochastic LLG equation can be found in \cite{brownthermal}, \cite{scholz2001micromagnetic}.The entire field acting on the nanomagnet $H_{eff}$ is given by,

\begin{equation}
\label{heff_eqn}
\vec{H}_{eff} = \vec{H}_{thermal}+\vec{H}_{aniso}+\vec{H}_{external}
\end{equation}
where $\vec{H}_{aniso}$ is the anisotropy field that, in in-plane magnets, is dominated by the demagnetization field arising from the shape of the magnet and is given by

\begin{equation}
\label{hdemag_eqn}
\vec{H}_{demag} = -M_S[N_{xx}m_x\widehat{x}, N_{yy}m_y\widehat{y}, N_{zz}m_z\widehat{z}]
\end{equation}
where $N_{xx}, N_{yy}, N_{zz}$ are the demagnetization factors that are calculated based on the analytical equations presented in Ref. \cite{aharoni1998demagnetizing}, and $m_x,m_y,m_z$ are the magnetization components of the nano-magnet in the $\widehat{x},\widehat{y}$ and $\widehat{z}$ directions. The presence of any external field can be included through the term $\vec{H}_{external}$.

\subsection{Stochasticity in Non-Telegraphic Regime}
The parallel and antiparallel states of the MTJ are stabilized by an energy barrier, $E_{B}$, that is defined as the product of the magnetic anisotropy and volume (Fig. \ref{enbar}). The retention time for the magnetic state of a nanomagnet is given by \cite{lopez2002transition},

\begin{equation}
\label{retention}
{T}_{retention} = {\tau}_{0}exp(\frac{E_B}{k_BT})
\end{equation}

where ${\tau}_{0}$ is a characteristic time constant in the range $1ps-100ps$ \cite{lopez2002transition}.The retention time or the lifetime of the magnet varies exponentially with the barrier height. The nonvolatility of the magnet enables such devices to be used in synchronous clocked systems where the device is operated in successive write and read phases. During the write cycle, a current pulse of fixed duration is passed through the HM layer that can switch the MTJ from one state over the barrier to the other stable state. The switching probability of the magnet varies with the magnitude of the current pulse flowing through the underlying HM layer. During the read phase, a small current is passed through the MTJ-${R}_{ref}$ (which can be implemented with another MTJ whose state is not disturbed by the small read current) voltage-divider circuit (see Fig. 2), and the MTJ state is read at the output of the inverter. The read current should be sufficiently small such that it does not disturb the state of the MTJ during the read phase. Since the voltage difference at the voltage-divider output for the parallel and antiparallel states is generally small, multiple stages of inverters are required to obtain a full swing at the output. 
\begin{table}
\centering
\caption{Device Parameters}
  \begin{tabular}{lSSSS}
\toprule\toprule
    \multirow{2}{*}{Parameter} &
      \multicolumn{4}{c}{Values} \\
      & {1$k_BT$} & {2$k_BT$} & {10$k_BT$} & {20$k_BT$} \\
      \midrule
   Free Layer Width, $W_{MTJ}$ & {10$nm$} & {17$nm$} & {30$nm$} & {40$nm$} \\
   Free Layer Length, $L_{MTJ}$ & {25$nm$} & {42.5$nm$} & {75$nm$} & {100$nm$} \\

   Free Layer thickness & \multicolumn{2}{c}{0.8 $nm$ } & \multicolumn{2}{c}{1.2 $nm$ }  \\
   Saturation magnetization, ${M_s}$ & \multicolumn{2}{c}{750 $KA/m$ } & \multicolumn{2}{c}{1000 $KA/m$ } \\
   Heavy metal thickness & \multicolumn{4}{c}{2$nm$} \\   
   Spin-Hall Angle, $\theta_{she}$  & \multicolumn{4}{c}{0.3 \cite{pai2012spin}} \\
   Gilbert's damping factor, $\alpha$ & \multicolumn{4}{c}{0.0122 \cite{pai2012spin}} \\
   Temperature, ${T}$ & \multicolumn{4}{c}{300$K$} \\

    \bottomrule
    \bottomrule
  \end{tabular}
\end{table}

Figure \ref{varyeb}(a) illustrates the variation of the MTJ switching probability with the amplitude of the current pulse being passed through the HM layer for different ${E}_{B}$ values. The device parameters used for simulations are enlisted in Table I. Note that the barrier height of the magnet is varied by scaling the area of the magnets appropriately. It can be shown that the probabilistic switching characteristics of the MTJ hold a sigmoidal relationship to the write current by describing the SHE layer current $I$, with two different parameters, namely $I_{bias}$ and $I_o$. $I_{bias}$ is the dc current required to bias the switching probability of the MTJ to 0.5, and $I_o$ is the scaling factor used to map the swing of the switching probability around the bias current to the sigmoid curve. Figure 4(b) depicts the variation of the switching probability of the MTJ with $I-I_{bias}$, normalized by a factor ${I}_{o}$. $I_o$ can be found by fitting the switching probability characteristics [$Psw(...)$] to the sigmoid function such that [refer Fig.\ref{varyeb} (b)],

\begin{equation}
\label{Io}
sigmoid( \frac{I-I_{bias}}{I_o}) \approx Psw(I)
\end{equation}

 As shown in Fig. \ref{varyeb}(a), when ${E}_{B}$ and hence, the device dimensions are scaled down, the current range required for stochastic switching decreases, thereby reducing the write current requirements of the device. Fig. \ref{varyeb}(c) indicates that both the components, $I_{bias}$ and $I_o$, are reduced with a reduction in barrier height. A reduction in $I_o$ implies that the current range that can be utilized for stochastic MTJ switching is reduced, thereby increasing the rate of change of the switching probability with a varying input current.
Consequently, the computing system becomes more prone to variations in the MTJ input current and exhibits less error resiliency with the reduction off $I_o$.  These considerations are highlighted in the next section.

Note that, if ${E}_{B}$ is not sufficiently large, the state of the magnet can switch during the read operation due to very small ${T}_{retention}$ value.  The retention failure probability ${P}_{F,retention}$, of a MTJ within a given read access time is given by

\begin{figure*}
\centering
\includegraphics[width=4.3in]{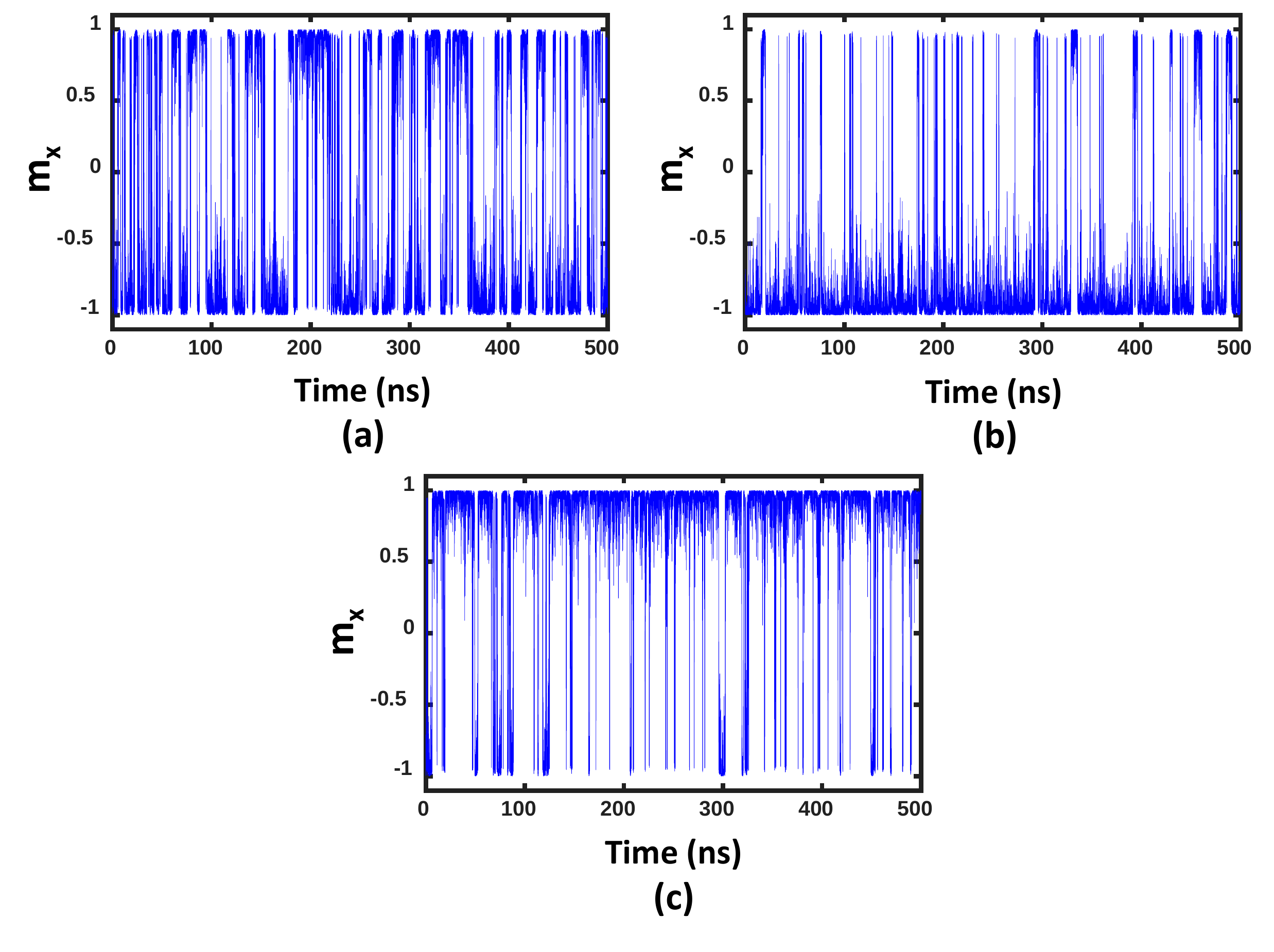}
\caption{\ Switching characteristics of an MTJ with 1${k}_{B}{T}$ barrier height: (a) When the current flowing through the HM is zero, the MTJ is equally likely to be in the parallel or anti-parallel state, (b) When $-1.5\mu A$ is flowing through the HM layer, the MTJ is more likely to be in the anti-parallel state, (c) When $1.5\mu A$ is flowing through the HM layer, the MTJ is more likely to be in the parallel state.}
\label{telegraphic}
\end{figure*}

\begin{equation}
\label{retention}
{P}_{F,retention} = 1-exp(-{t}_{read}/exp(\Delta))
\end{equation}

where ${P}_{F,retention}$ is the retention failure probability of the MTJ during a read time of ${t}_{read}$ in nano-seconds, and $\Delta$ is the ${E}_{B}$ of the MTJ in $k_{B}T$. In order to find the necessary ${t}_{read}$ for correct read operation, SPICE simulations (with a Verilog A model for the MTJ \cite{fong2011knack}) are performed in IBM $45nm$ technology node. Simulation results show that the required read time is around $0.2ns$ for the nominal corner and $1ns$ for the worst case corner (with 2$\sigma$ variations in the threshold voltage of the CMOS transistors). Hence, for retention failure probability calculations, the required read time is taken to be $1ns$ to ensure that a correct read can be achieved even for the worst corner. As illustrated in Fig. \ref{varyeb}(d), retention failure probability increases exponentially as the MTJ is scaled down. In order to keep the retention failure probability smaller than $1\%$, the ${E}_{B}$ of the magnet should be kept greater than $4.6k_{B}T$. When the MTJs are scaled further they enter the superparamagnetic regime, where the magnets are no longer thermally stable during the read cycle. Hence, parallel read-write operations are required for magnets in the superparamagnetic regime ($E_{B}<5k_{B}T$) to realize stochastic switching elements.    

\subsection{Stochasticity in the Telegraphic Regime}

For low barrier height nanomagnets ($E_{B}\sim1k_BT$), even with zero charge current flowing through the HM layer, the MTJ exhibits random telegraphic switching between the two equilibrium states (Fig. \ref{telegraphic}(a)) due to thermal noise. The random switching characteristics of such scaled devices in the superparamagnetic regime can be still manipulated by passing a current through thr HM layer. For instance, Figs. \ref{telegraphic}(a)-(c) represents the in-plane magnetization of the MTJ in presence of write current of $0, 1.5,-1.5 \mu A$, respectively, flowing through the HM layer of a $1k_{B}T$ magnet. The dwell time of the MTJ in either of the two stable states can be modulated by the magnitude and direction of the input write current.

\begin{figure*}[t!]
\centering
\includegraphics[width=4.4in]{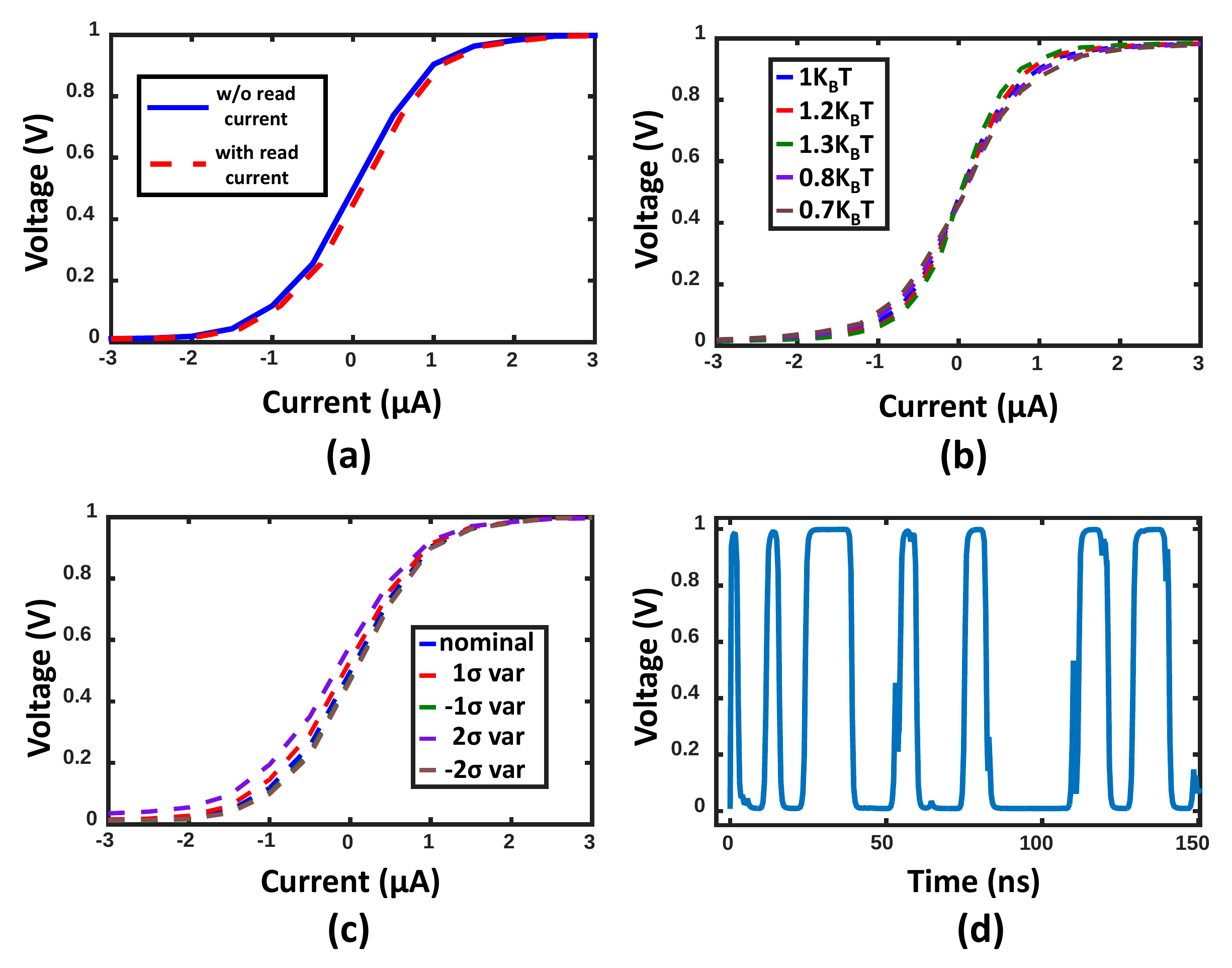}
\caption{\ (a) Average inverter output over a duration of $2\mu s$ with and without the impact of the read current, (b) Variation of the inverter average output  over a duration of $2\mu s$ with magnitude of the write current for different ${E}_{B}$ values, (c) Inverter average output  over a duration of $2\mu s$ for nominal corner and for the worst case conditions of $\pm$1$\sigma$ and $\pm$ 2$\sigma$ variations in the threshold voltages of the transistors, (d) A typical plot of the output voltage of the inverter stage of the read circuit as a function of time under zero external input current. }
\label{devvar}
\end{figure*}

The volatility of these devices entails a rethinking of the manner in which such nanomagnets can be operated with peripherals to realize a stochastic computing element. Because of device volatility and low retention time, such devices cannot be operated with separate write and read phases. Consequently, the write and read terminals of the MTJ are activated simultaneously, and the device state is read while an input bias current flows through the underlying HM layer of the MTJ. For high-energy-barrier MTJs, the effect of the read current on the switching characteristics is not a design issue since the read and write cycles are decoupled in time. However, for MTJs in the telegraphic switching regime, the read current can bias the switching characteristics since the read and write operations occur in parallel. Furthermore, since the devices are highly scaled, the write (for stochastic switching) and read currents fall in the same order of magnitude (unlike high-barrier-height magnets, where the write current for stochastic switching is higher). Hence, the resistive divider of the read circuit (Fig. \ref{shemtj2} )needs to be highly optimized such that the read current is maintained at the minimal value. SPICE simulations reveal that the read current can be minimized to 100 nA while having a minimal effect on the MTJ switching characteristics. Figure \ref{devvar}(a) depicts the average output of the inverter stage over a duration of $2\mu s$ with and without the read current. The case ``with read current" is simulated by considering the additional spin-orbit torque induced by the $100nA$ read current flowing through the HM layer  while the case ``without read current" ignores the effect of the additional read current. As can be observed in Fig. \ref{devvar}, the read current has minimal impact on the MTJ switching probability. Furthermore, device dimension variations (or equivalently $E_B$ variations) and read circuit variations ($\pm$1$\sigma$ and $\pm$ 2$\sigma$ variations in the threshold voltages of the CMOS transistors) are shown to have minimal effect on the stochastic switching behavior of the nanomagnets (Figs. \ref{devvar}(b)-(c)). Figure \ref{devvar}(d) represents a typical plot of the voltage output of the inverter stage as a function of time with no input current flowing through the underlying HM of the MTJ.

Note that the switching characteristics of superparamagnetic MTJs are highly sensitive to any change in the magnitude of the write current. As depicted in Fig. \ref{devvar}(a), the switching probability of the MTJ shifts from 0.5 to 0.85 for a $1\mu A$ change in the write current. Hence, the impact of variations on the input current provided to a network of such scaled MTJs can be significant, and it is analyzed in more detail in the next section. We would like to conclude this section by mentioning that the parallel read-write operation is not suited for magnetization switching in the nontelegraphic regime [$(10-20)k_BT$ barrier height magnets] since the telegraphic switching would occur in time scales ranging approximately from micro- to milliseconds, thereby resulting in an enhanced delay for the computing process.

\begin{figure}[t!]
\centering
\includegraphics[width=3.2in]{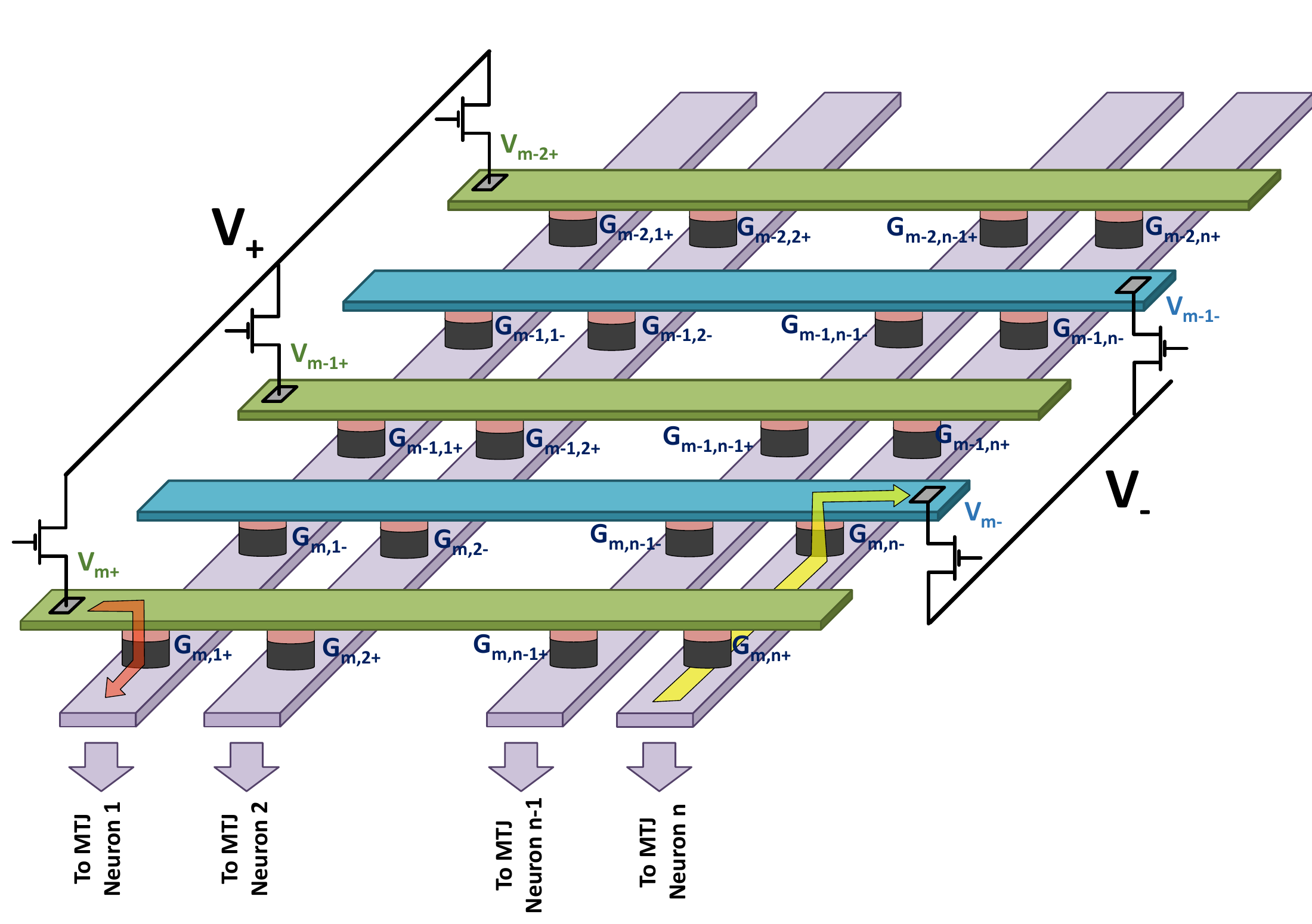}
\caption{\ Crossbar architecture connecting the inputs of one layer to the neurons of the corresponding layer. Horizontal bars provide the input voltage for the synapses. The summation of weighted synaptic currents along the columns of the crossbar array are then provided as inputs to the MTJ neurons.}
\label{crossbar}
\end{figure}

\subsection{Stochastic Neuromorphic Computing}

A neural network is essentially a collection of layers of neurons interfaced through a network of weighted synapses. A particular input to a neuron is first scaled by the corresponding synaptic weight of the synapse before it is accumulated and processed by the neuron. Neurons with sigmoid like transfer functions have been shown to be have appeal for implementing deep spiking neural networks \cite{probabilistic}, making SHE-MTJ structures ideal for realizing energy-efficient neuromorphic hardware. In the stochastic neural network considered in this work, the MTJ neuron generates an output spike probabilistically depending on the instantaneous magnitude of the resultant weighted synaptic input \cite{probabilistic}. This computing framework can be directly translated to the resistive crossbar architecture illustrated in Fig. \ref{crossbar}, where the synaptic weights are mapped into the resistive elements between the horizontal and vertical metal lines.

 Note that resistive crossbar arrays based on memristive devices like phase-change materials \cite{kuzum2011nanoelectronic}, Ag-Si devices \cite{jo2010nanoscale} and spintronic devices \cite{sengupta2016hybrid} have been proposed and experimentally demonstrated \cite{prezioso2015training}. Two horizontal lines are used for each input connected to the crossbar array to implement the functionality of positive and negative weights. An input spike provided to the network activates the corresponding access transistors by supplying a voltage to the horizontal lines ${V}_{+}$ (positive voltage) and ${V}_{-}$ (negative voltage), which is translated to a current through the vertical columns (weighted by the conductances of the resistive elements). The current accumulated in the vertical columns are then supplied as the write currents to the stochastic neurons of the corresponding layer. 
 
 If the weight connecting an input $m$ to a neuron $n$ is negative, then the corresponding resistive element connecting the positive horizontal line and the vertical column (${G}_{m,n+}$ ) is programed to a high resistive \textit{off} state and the weight connecting the vertical column and the negative horizontal line is programed to a conductance given by ${G}_{m,n-} = {w}_{m,n}{G}_{o}$ and vice versa.  Here, ${w}_{m,n}$ is the synaptic weight between the input $m$ and neuron $n$ and ${G}_{o}$ is the mapped conductance for unity weight. The conductances of the resistive elements are selected by scaling the synaptic weights by a factor ${G}_{o}$ given by, [${I_o}/{(\delta V)}$],  where ${\delta V}$ is the magnitude of the supply voltage driving the rows of the crossbar array and ${I_o}$ is the current scaling factor of the stochastic MTJ mentioned previously. Assuming that the magnetometallic spin devices have low input resistance compared to the cross-point resistances of the crossbar array, the neurons receive a weighted summation of spike inputs in a particular layer and produce output spikes probabilistically over time that will drive the fan-out neurons of the next layer. For magnetic neurons operating in the nontelegraphic regime, the read circuit can be interfaced with a latch that stores the inverter output during the read cycle, which drives the next stage of neurons during the following write cycle (hence the term synchronous operation).
 
For magnetic neurons operating in superparamagnetic regime, the inverter output can directly drive the neurons in the next stage (hence asynchronous operation).  Note that the high-barrier-height magnets are also driven by a current source to bias it at a switching probability of 0.5, unlike MTJs in the superparamagnetic regime. Owing to the small input current and the zero bias current of magnetic neurons operating in the superparamagnetic regime, asynchronous architectures grant significant power savings in the neurons and the resistive crossbar
array. However, as shown later, asynchronous implementation incurs significant power loss at the read circuit owing to the continuous switching activity of the inverters.

\section{Design Considerations: Synchronous and Asynchronous Neuromorphic Systems}

\subsection{Device to System Simulation Framework}
In order to analyze the design considerations for synchronous and asynchronous stochastic SNNs, a hybrid device-circuit-system cosimulation framework is used in this work. A stochastic LLGS simulation for MTJs with different barrier heights is used to evaluate the probabilistic switching behavior of magnets operating in the nontelegraphic to telegraphic regime. In this work, we use magnets having barrier heights $10$ and $20k_BT$ for nontelegraphic regime and magnets of barrier height $1$ and $2k_BT$ for telegraphic regime. The device parameters used for simulations are summarized in table I. SPICE-level simulations based on a Verilog-A model of the MTJ is used to evaluate the performance of the stochastic MTJ along with associated peripherals.

\begin{figure}[t!]
\centering
\includegraphics[width=2.8in]{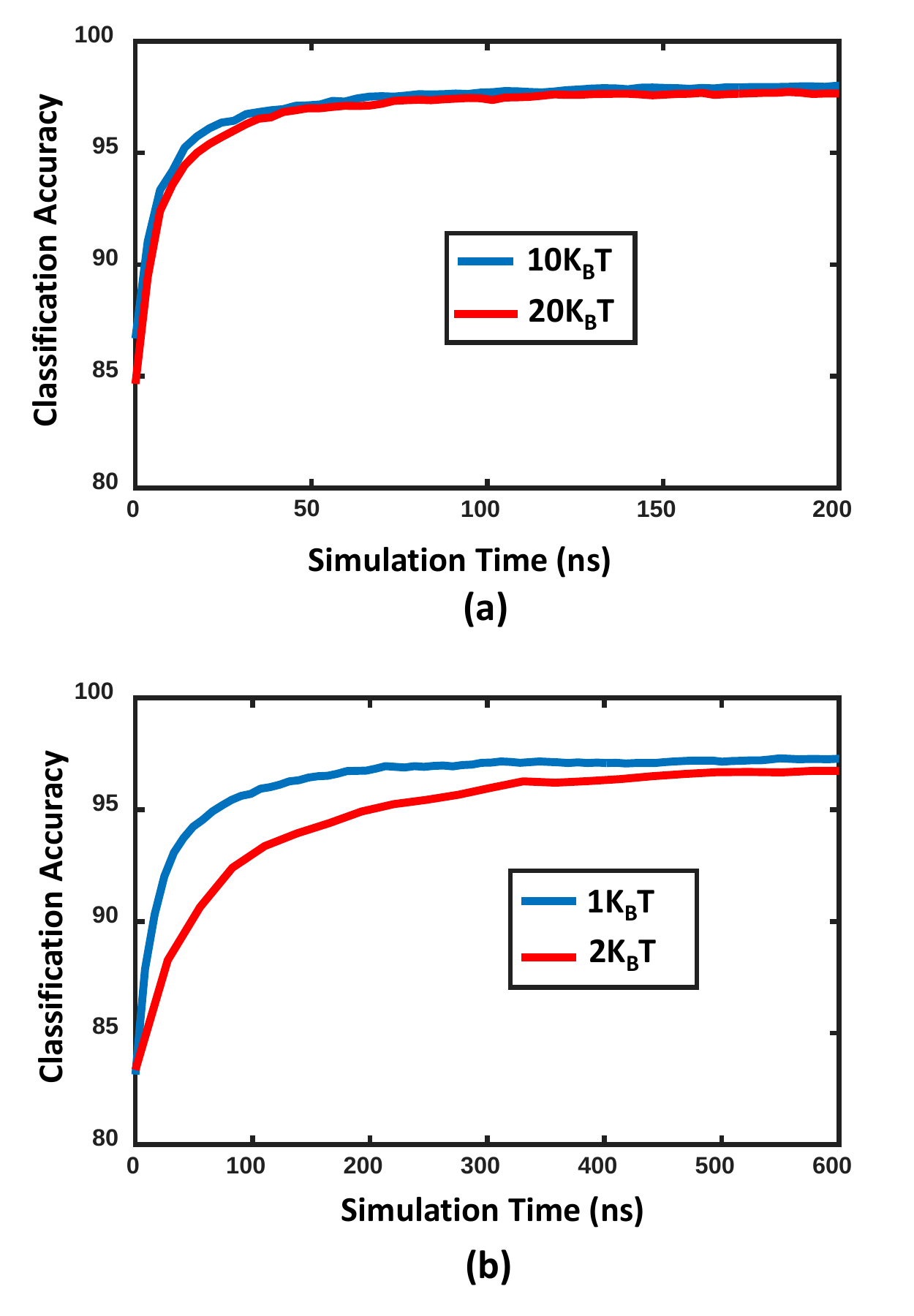}
\caption{\ Variation of classification accuracy of the proposed network with time for (a) Synchronous, and (b) Asynchronous implementations.}
\label{accuracy}
\end{figure}

\begin{figure*}[t!]
\centering
\includegraphics[width=4.7in]{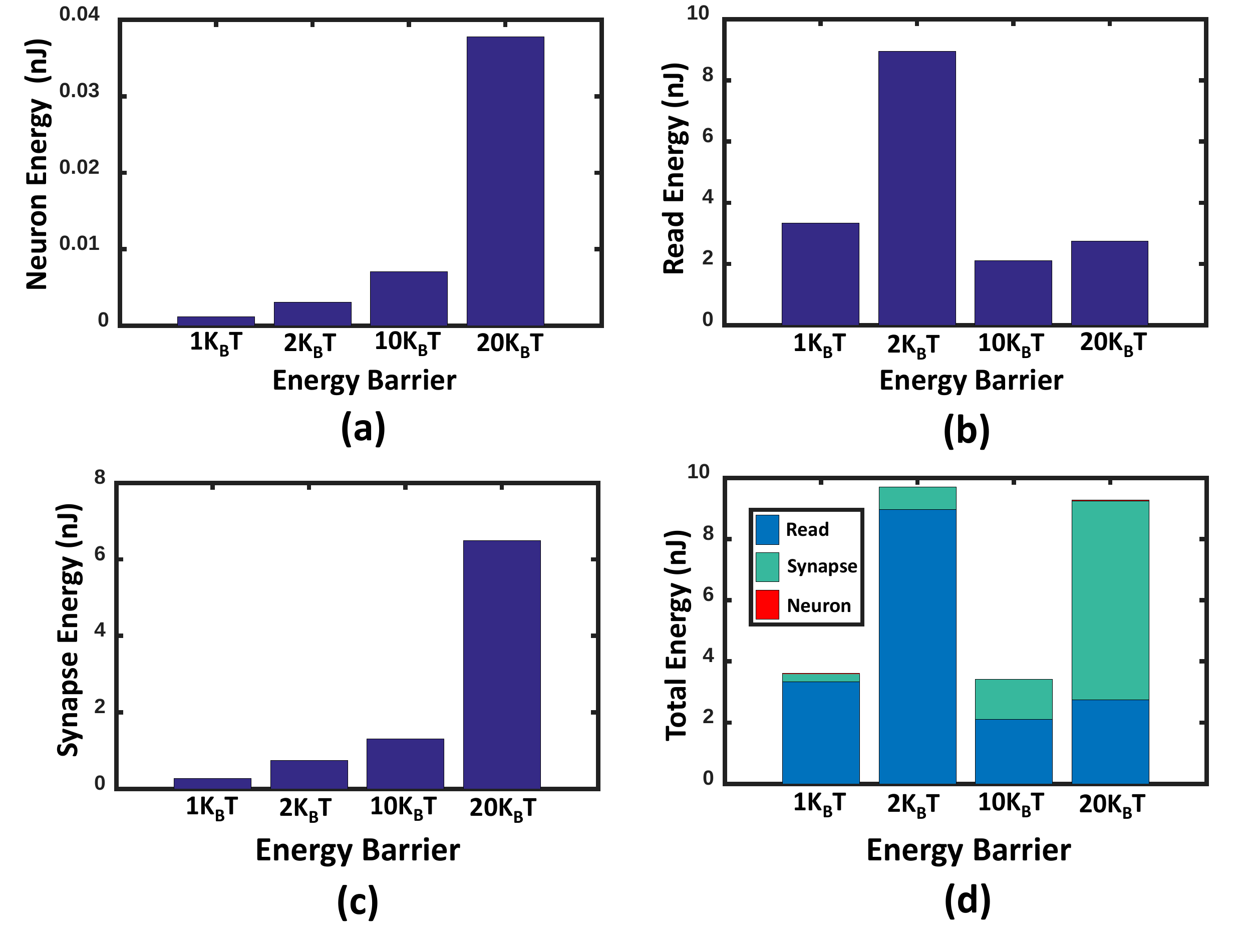}
\caption{\ (a)  Energy consumption of the MTJ neuron, (b) Energy consumption of the read circuit, (c) Energy consumption of the synapses, (d) Total energy consumption per image classification (for an accuracy of 96$\%$) for the asynchronous (1${k_BT}$ $\&$ 2${k_BT}$) and synchronous (10${k_BT}$ $\&$ 20${k_BT}$) networks. }
\label{power}
\end{figure*}

In order to perform a system-level analysis, the performance of the network is assessed for a large-scale deep-learning network architecture (28$\times$28-6c5-2s-12c5-2s-10o) on a standard digit-recognition problem based on the MNIST dataset \cite{palm2012prediction}. The network consists of alternate layers of convolutional and subsampling operations. The dimensions of the input MNIST images are 28$\times$28, which are applied as input to the convolutional layer consisting of six convolutional kernels with a size of 5$\times$5. The subsampling kernel has a size of 2$\times$2 and is followed by another convolutional layer comprising of 12 output maps, which, in turn, is followed by another subsampling layer. The final layer consists of ten neurons, each of which represents one of the ten digit classes. Once the training is accomplished, the learned weights are mapped to the synaptic conductances using a value of $G_{o}= 5\mu S$ which is in the typical resistance range for memristive synaptic devices. The same resistive crossbar array is used for all of the different barrier-height neuronal devices. The supply voltage $\delta V$ was adjusted in each case to satisfy the relationship, $\delta V={I_o/G_o}$, as explained previously. The supply voltages $\delta V$, was calculated to be $0.1$,$0.11$,$1.05$ and $2V$ for nano-magnets of barrier height $1$,$2$,$10$ and $20k_BT$ respectively. The sigmoid-curve characteristics for the magnets operating in the telegraphic regime are obtained by averaging the output voltage of the read inverter circuit over a period of $2\mu s$ (for $1k_BT$ ) and $5\mu s$ (for $2k_BT$).

\begin{figure*}[t!]
\centering
\includegraphics[width=4.7in]{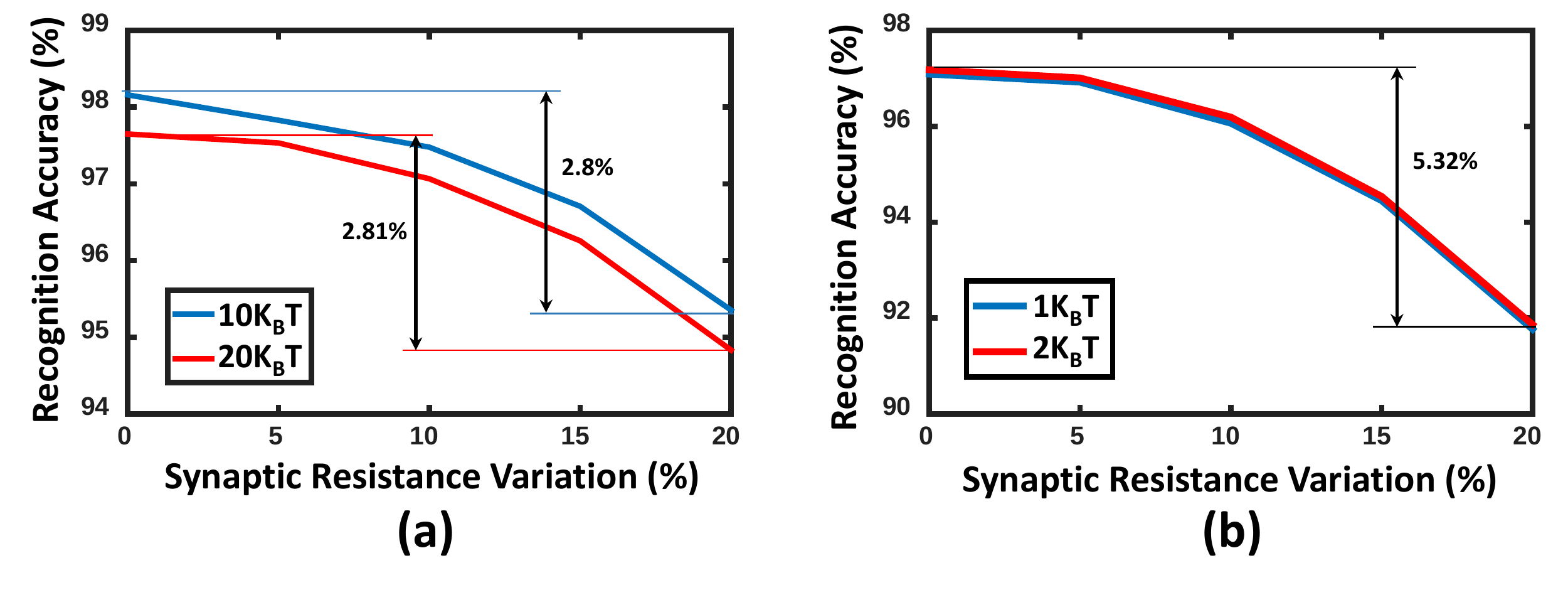}
\caption{\ Average classification accuracy (measured over 50 independent Monte Carlo simulations) with variations in the resistive synapses ($\%\sigma$ variations) for the (a) synchronous design, (b) asynchronous designs.}
\label{vary_cb}
\end{figure*}

\subsection{Performance and Energy Estimation}
Figure \ref{accuracy} depicts the temporal evolution of the classification accuracy of the stochastic SNN for the synchronous and asynchronous designs. For the 10 and the 20$k_BT$ synchronous designs the classification accuracy reaches 98.1\% and 97.6\% respectively, while it saturates at 97.5\% and 97.2\% for the 1 and 2$k_BT$ asynchronous designs. Both synchronous networks surpass an accuracy of 95\% just under 20ns, whereas the two asynchronous networks require 80ns (for 1$k_BT$) and 250ns (for 2$k_BT$) to reach the same accuracy. In the asynchronous implementation, the high frequency telegraphic switching of the nano-magnets is translated into voltage spikes at a lower frequency due to gate capacitance charge delays of the CMOS devices, which explains the slower response of the asynchronous networks compared to the synchronous designs. Also as the $E_B$ values of the nanomagnets are increased (for the superparamagnetic regime), the retention time of the nanomagnets increase, decreasing the spiking frequency at the output of the inverters. Hence, as the results show, for asynchronous designs, the time required for a network to reach a target accuracy increases with the $E_B$ value of the nanomagnets used in the design. For the synchronous
networks, the duration of one time step is selected to be 4 ns, which includes a write time of 0.5 ns, a rest period of 2 ns, and a read time of 1 ns, followed by a reset period of 0.5 ns. The duration of the time step for the asynchronous networks is determined by measuring the average duration of a voltage pulse at the output of the inverter read circuit at zero write current, and it is calculated to be $8.2$ and $27.5ns$  for the $1$ and $2k_BT$ networks, respectively.

\begin{figure*}
\centering
\includegraphics[width=4.4in]{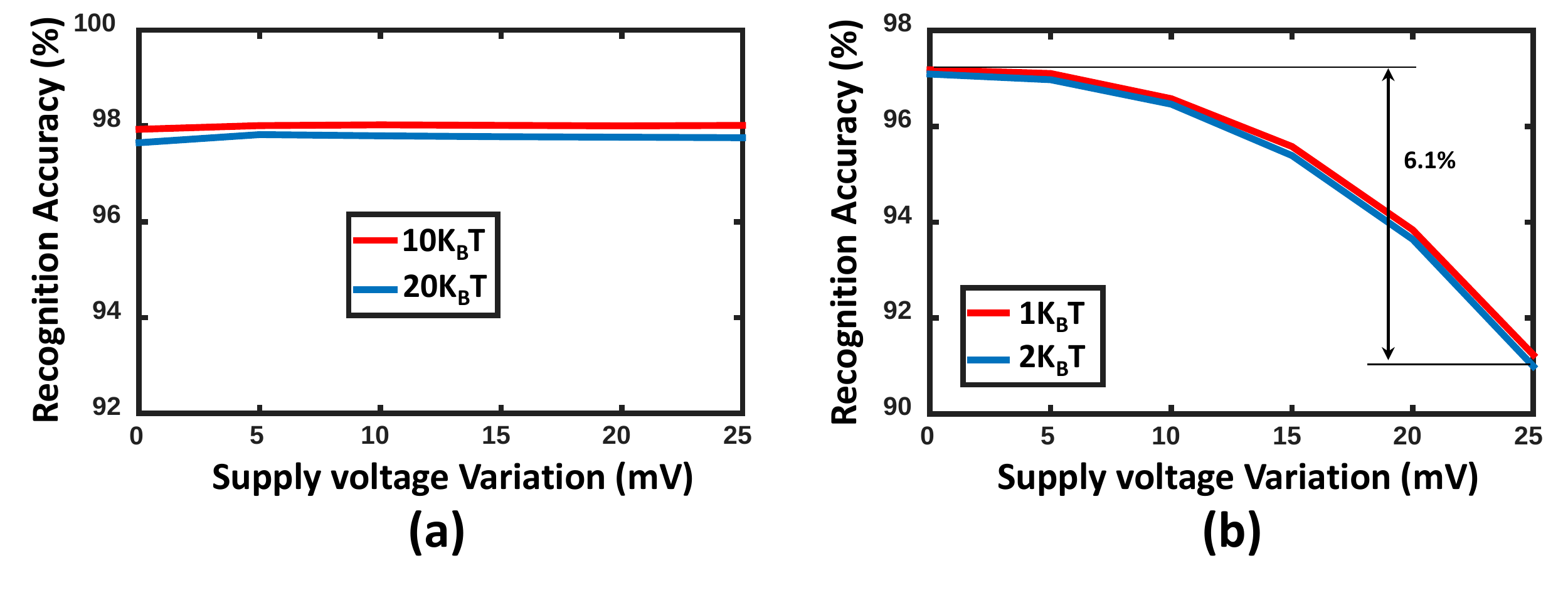}
\caption{\ Average classification accuracy (measured over 50 independent Monte Carlo simulations) with variations in the supply voltage (up to 25$mV$ variations) for the (a) synchronous, and (b) asynchronous designs.}
\label{vary_v}
\end{figure*}

Figure \ref{power} summarizes the energy consumption observed for different components of the network (both synchronous and asynchronous) corresponding to a target classification accuracy of 96\%. Neuron energy [Fig. \ref{power}(a)] refers to the energy dissipated in the MTJ neuron due to the write-reset currents flowing through the HMlayer. The neuron energy consumption is lowest for the 1$k_BT$ asynchronous design with an energy consumption of 1.15$pJ$ per image classification, and increases with the size of the magnets up to 37.8$pJ$ per image classification for the 20$k_BT$ synchronous design. This trend can be explained by the increasing write-current requirements of the nanomagnets as their sizes are increased. Since the current flowing through the HM layer is first routed through the resistive crossbar network (the synapses), the energy consumption in the synapses [Fig. \ref{power}(c)]  show a similar trend, increasing with the size of the magnets. Also, the bias current required in the synchronous designs to bias the switching probability of the MTJs to 0.5 adds to the power dissipation in the HM layer and the synapses. The energy-consumption values in the synapses per image classification are 0.27 and 0.74$nJ$ for the 1 and 2$k_BT$ asynchronous designs, and 1.3 and 6.5$nJ$ for the 10 and 20$k_BT$ synchronous designs. The read energy consumption, illustrated in Fig. \ref{power}(b), is the summation of the power dissipated in the MTJ due to the read current passing through and the power dissipated in the CMOS interface circuitry. As the results indicate, the read energy consumption per image classification are larger for the asynchronous implementations (3.3 $nJ$ for the 1$k_BT$ and 8.95$nJ$ for the 2$k_BT$)  than for the synchronous implementations (2.1$nJ$ for the 10$k_BT$ and 2.75$nJ$ for the 20$k_BT$). The majority of the read power dissipation in asynchronous networks occur at the CMOS inverters, which are required to operate continuously due to the parallel read-write nature of the neurons. In synchronous networks, however, the CMOS inverters are required to operate only during the read cycle, and can be deactivated at other times using access transistors to save power. For both designs the power dissipated in the neurons are an order of magnitude smaller than the power dissipated in the synapses and the read circuit, owing to the low resistance of the HM layer. As depicted in Fig. \ref{power}(d), the 10$k_BT$ synchronous network shows the minimum energy requirement per image classification (3.4$nJ$), closely followed by the 1$k_BT$ asynchronous network  (3.6$nJ$). The 2$k_BT$ asynchronous network exhibit an energy consumption of  9.7$nJ$ per image classification followed by the 20$k_BT$ synchronous network with an energy consumption of 9.28$nJ$. For the synchronous networks, the energy consumption associated with the clocking circuitry is negligible, especially since a classification accuracy of 96\% can be achieved under 10 clock cycles, and hence is not considered in this analysis. 
\\

\subsection{Effect of Variations}
Most of the computations of the proposed network occur in the resistive crossbar array. Hence, any variations in the resistive elements of the crossbar array can result in a significant degradation of the classification accuracy. To measure the effect of such variations, separate experiments are performed allowing variations with a standard deviation up to 20\%in the resistive elements. According to the results (see Fig. \ref{vary_cb}) for variations in the synapses with a standard deviation of 20\%, the accuracy loss is only 2.8\% for the synchronous designs and 5.32\% for the asynchronous designs. The slightly higher accuracy degradation observed in the asynchronous designs in comparison to the synchronous designs can be explained by the increased sensitivity of the MTJ switching probability in response to the write current at the superparamagnetic regime.

Because of the low operating currents of the nanomagnets used in the asynchronous design, the operating voltage of the crossbar architecture given by $\delta V={I_o}/{G_o}$ can be very small for low $k_BT$ magnets. Hence any variation in the supply voltage can potentially result in a large deviation in the write-current magnitude, influencing the classification accuracy of the network. Figure  \ref{vary_v} depicts the behavior of the classification accuracy of the two designs in the presence of supply voltage variation. As shown in Fig. \ref{vary_v}(a), owing to the larger supply voltages used in the synchronous designs, 10 and 20$k_BT$ synchronous implementations are resilient to supply voltage variations up to 25$mV$. The asynchronous implementations, on the order hand, exhibit an accuracy degradation of 6.1\% when variation in the supply voltage is less than 25$mV$. 

\begin{figure}[b!]
\centering
\includegraphics[width=2in]{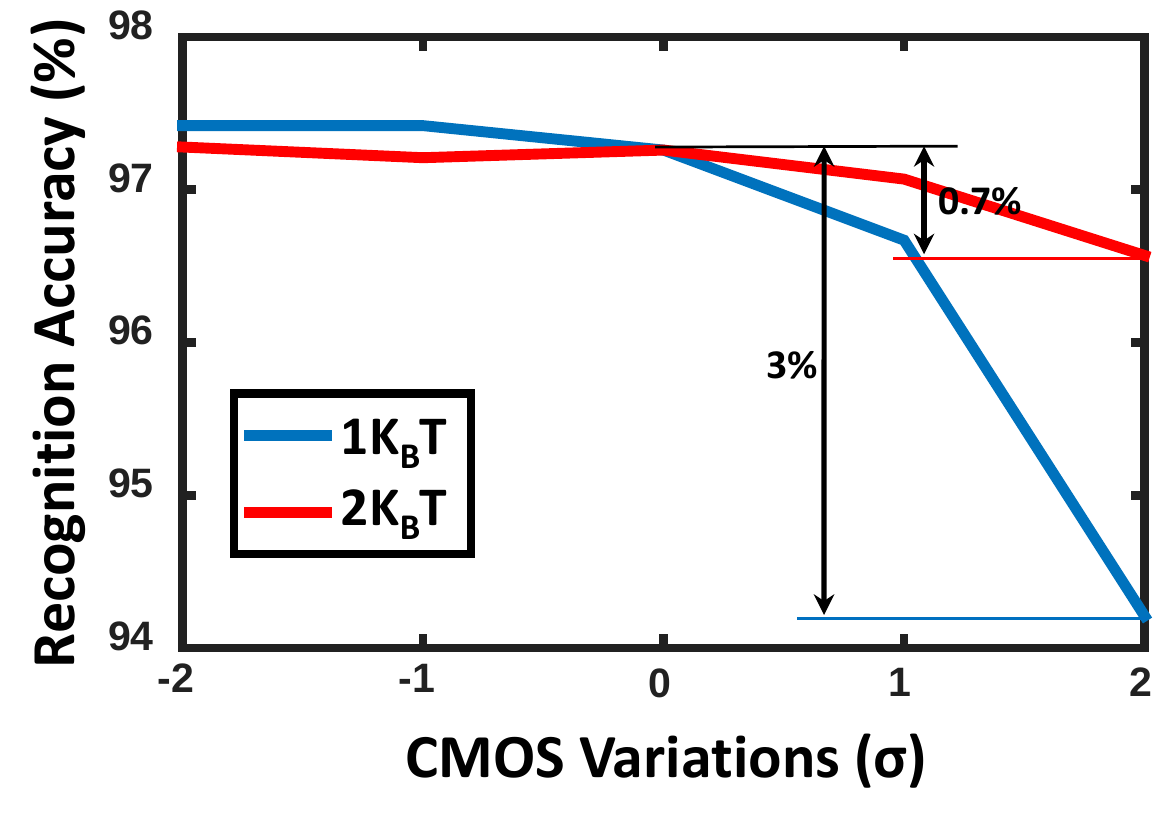}
\caption{\ Average classification accuracy for the worst case corner, with variations in the CMOS read circuit (upto $\pm$2$\sigma$ variation) for the asynchronous design. }
\label{vary_cmos}
\end{figure}

As explained in Sec. II, the CMOS inverter read circuit for the asynchronous implementation must be designed carefully so that the average magnetization of the nano- magnet is properly reflected on the average output of the inverter. Any variation in the CMOS circuitry can offset the average output of the inverters, adversely affecting the classification accuracy of the network. As depicted by Fig. \ref{vary_cmos}, the classification accuracy of the 1$k_BT$ asynchronous network decrease by 3$\%$ and the accuracy of the 2$k_BT$ asynchronous network decrease by 0.7$\%$ for the worst case corner with 2$\sigma$ variations in the CMOS read circuit. The synchronous networks are resilient towards such CMOS variations since the read time is selected to be adequate for a correct read even at the worst cell corner. 

\begin{figure*}
\centering
\includegraphics[width=4.4in]{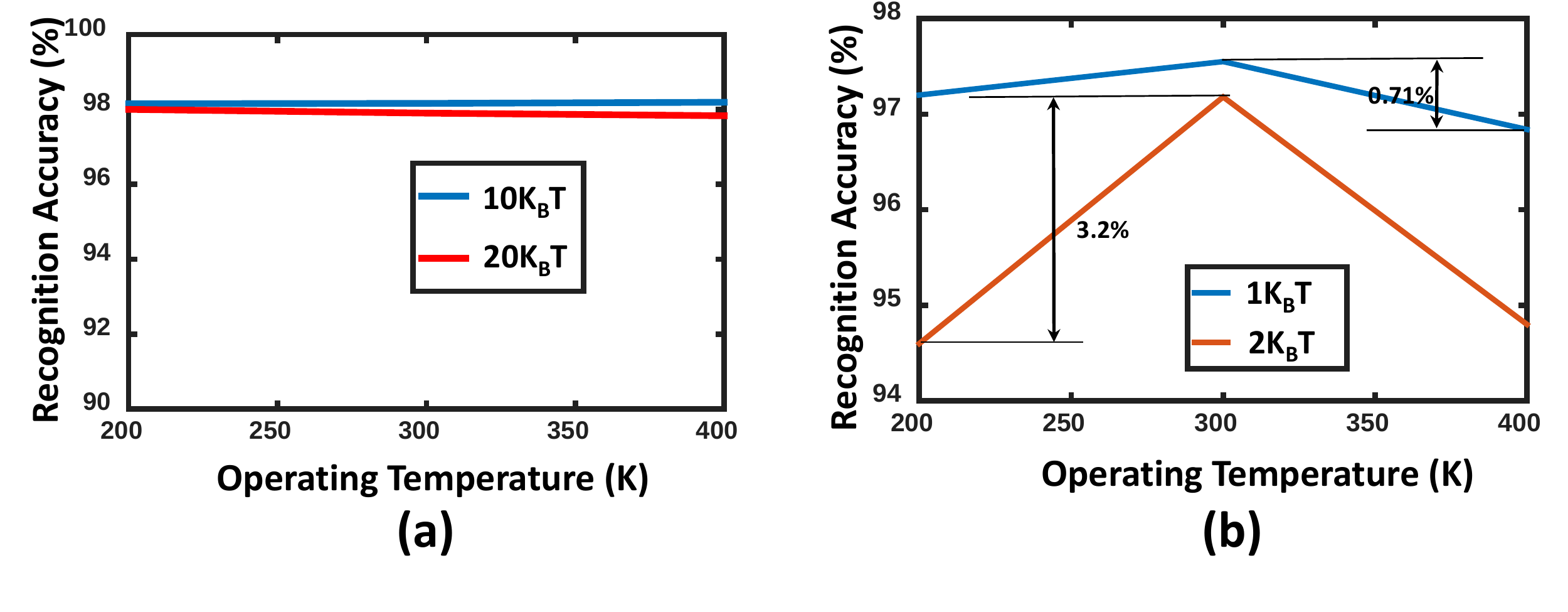}
\caption{\ Classification accuracy with varying operating temperature for the (a) synchronous, and (b) asynchronous designs.}
\label{vary_temp}
\end{figure*}

\begin{figure*}
\centering
\includegraphics[width=4.4in]{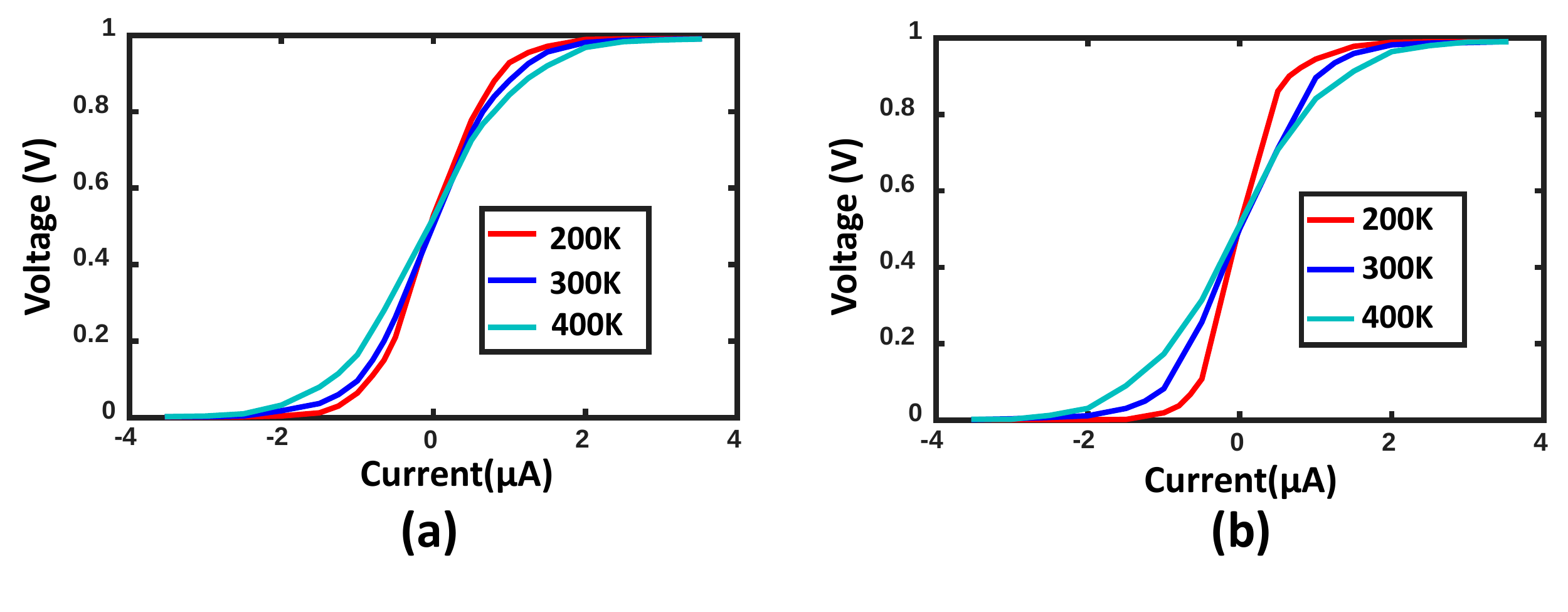}
\caption{\ Average inverter output under different temperatures for (a) $1k_BT$ magnet   (b) $2k_BT$ magnet .}
\label{sw_temp}
\end{figure*}

\subsection{Effect of Temperature}
In this work, the switching characteristics of the MTJs are varied between the telegraphic and nontelegraphic regimes by adjusting the width of the FL appropriately. However, the switching characteristics of the MTJs can deviate significantly from design values as the operating temperature changes. Figure \ref{vary_temp}) depicts how the classification accuracy of the two designs vary as the operating temperatures are changed from 200 to 400 K. As observed by the simulation results, the two synchronous networks are resilient to variations in temperature and show an error degradation of less than 0.4\% at 400 K. The two asynhronous networks, on the other hand, are not as resilient to variations in temperature. The $1k_BT$ network display an accuracy degradation of $0.71\%$ at 400K and $0.6\%$ at 200K, while the $2k_BT$ network display an accuracy degradation of $2.8\%$ at 400K and $3.2\%$ at 200K. The higher temperature dependency of the $2k_BT$ network  can be explained by the change in the switching characteristics of the MTJs at different temperatures. As illustrated by  Fig. \ref{sw_temp}) the average inverter output of the $2k_BT$ magnet displays a larger shift with temperature than the $1k_BT$ magnet, resulting in a higher accuracy degradation.

\section{Summary}

In this paper, we outline the design considerations for MTJ-based stochastic SNNs with varying barrier heights. We show that the reduced energy consumption of low-barrier-height magnets is achieved at the expense of reduced error and variation tolerance and constrained design space for the CMOS peripherals. We further show that, contrary to the popular belief that superparamagnetic MTJs are more energy efficient than high-barrier-height magnets, parallel and always on ``read" and ``write" operations in superparamagnets cause the peripheral read
circuit energy consumption to dominate the network energy-consumption profile. While scaling in the peripheral CMOS technology reduces the peripheral energy consumption, reduced error tolerance might still be a concern for spin-based neuromorphic hardware design. The analysis performed in this work can be easily extended to other applications that require probabilistic inference—for example, Bayesian networks and Ising computing.
\newline


\section*{Acknowledgment}

The work was supported, in part, by the Center for Spintronic Materials, Interfaces, and Novel Architectures (C-SPIN), a MARCO- and DARPA-sponsored StarNet center, by the Semiconductor Research Corporation (SRC), the National Science Foundation (NSF), Intel Corporation, and the U.S. DoD Vannevar Bush Fellowship.
\bibliography{Main_text_incl_figures}

\end{document}